# DESIGNING MID-AIR HAPTIC GESTURE CONTROLLED USER INTERFACES FOR CARS

Gareth Young, Hamish Milne, Daniel Griffiths, Elliot Padfield, Robert Blenkinsopp, and Orestis Georgiou

Ultraleap Ltd.  The West Wing, Glass Wharf, Bristol, BS2 0EL, United Kingdom

We present advancements in the design and development of in-vehicle infotainment systems that utilize gesture input and ultrasonic mid-air haptic feedback. Such systems employ state-of-the-art hand tracking technology and novel haptic feedback technology and promise to reduce driver distraction while performing a secondary task therefore cutting the risk of road accidents. In this paper, we document design process considerations during the development of a mid-air haptic gesture-enabled user interface for human-vehicle-interactions. This includes an online survey, business development insights, background research, and an agile framework component with three prototype iterations and user-testing on a simplified driving simulator. We report on the reasoning that led to the convergence of the chosen gesture-input and haptic-feedback sets used in the final prototype, discuss the lessons learned, and give hints and tips that act as design guidelines for future research and development of this technology in cars.

## 1   INTRODUCTION

At the dawn of the age of autonomous vehicles, alarming reports are showing that every year approximately 26,000 people die, and 1.4 million people are injured in traffic accidents all around Europe [1]. The great majority of these road accidents (about 90%) involve human error [2]. Specifically, distracted driving in the USA accounts for approximately 25% of all motor vehicle crash fatalities, thus contributing to 3,477 traffic deaths in 2015 alone [3]. These reports and findings provide strong arguments in favor of driver automation which promises to significantly reduce these figures [4]. Meanwhile, and despite the clear indications that more care is needed when designing human machine interfaces (HMIs) to be operated by drivers, approximately 50 million cars were equipped with touchscreens in 2017, an interface that is by its very nature visually distracting. Therefore, until autonomous vehicles become widespread, much effort is still needed to improve HMIs in cars that ensure driver safety without compromising user experience (UX) and design.

Already, car manufacturers such as BMW, Mercedes, VW, Cadillac, Jaguar, and Hyundai see great potential in mid-air gesture interfaces [5]. Mid-air gestures are generally facilitated by optical tracking systems that apply machine vision algorithms to recognize hand gestures (HGs) that are translated into input commands or responses to the in-vehicle infotainment system (IVIS) functions. A key advantage of mid-air HG interaction interfaces is their ability to reduce the mental and visual demand of the IVIS on the driver [6]-[10]. Meanwhile, the primary disadvantage of pure HG interaction control is the loss of tactile feedback, a key ingredient towards the sense of agency – the subjective experience of voluntary control over your actions. This loss of physicality in turn also lessens user satisfaction, safety and usability of HG interactions since the user is often unaware if their input HG command has been correctly read by the car computer. Other input modalities such as speech and non-speech audio feedback have been proposed to offset this lack of tactile feedback [9], however, these methods are not as effective as one would like and are subject to interference from other audio signals and external noise, e.g., due to an open window, or an ongoing conversation in the car.

To that end, a promising solution set that can re-instill physicality and the sense of touch to HGs is that of mid-air haptics. Including tactile feedback to HGs could for example reduce the screen-looking time [18] [19], button locating time, improve ergonomics, selection accuracy, the sense of agency [11] and UX altogether, while not disrupting ongoing conversations or disturbing other passengers. Technologies capable of delivering such tactile solutions include, for example, mid-air focused ultrasonic haptic interfaces [12]-[15], air jets and cannons [16]. With the exception of focused ultrasound haptic feedback, the other technologies currently exist only in research laboratories.

Focused ultrasound haptics was invented in Japan in 2010 [13] and commercialized in 2013 by Ultrahaptics (now Ultraleap) in the UK [14]. Using an array of ultrasound speakers, tactile sensations are projected onto the user's hands and fingers for example to indicate acknowledgement of a certain HG input command. Such a system has been studied extensively in the literature, both with regards to its hardware architectures and software algorithms and is now being tested in various application settings including digital signage [17] and mid-air interactive IVISs found in cars [18-20]. Indeed, initial studies at Nottingham [19] and Glasgow [18] universities suggest that combining ultrasonic mid-air haptic sensations with intuitive HGs has the potential to eliminate the need for vision while interacting with the IVIS completely. For example, discrete virtual buttons could be presented in three-dimensional space to replicate the familiar layout of a traditional touchscreen interface (i.e., as an ordered array of buttons). Once identified, selection could be made by physically moving the hand downwards (to emulate the pressing a virtual button). Additionally, the virtual array of targets can be placed anywhere in three-dimensional space in proximity to the steering wheel and can even be adjusted to the user's preferred height or preferred reaching distance (akin to a height-adjustable steering wheel) and is therefore not bound to an existing surface, infrastructure, or car interior design [19]. Indeed, further studies in Palermo have demonstrated that it is possible to recognize and navigate a purely tactile version of a hierarchical in-air icon-based menu that is generated by focused ultrasound technology [43] using HGs and mid-air haptics, without using any visual or verbal inputs.

In an automotive setting, this level of flexibility and customization offers ample benefits in terms of anthropometrics and physical ergonomics, as well as potential space savings in vehicles, however, also presents new and largely unexplored challenges to the automotive user interface (UI) designer and researcher [21]. *How should one design the IVIS interface as to maximize on the afforded benefits of mid-air haptic HGs? Conversely, how should one design said interface to avoid any potential shortcomings or pitfalls of this new technology?*

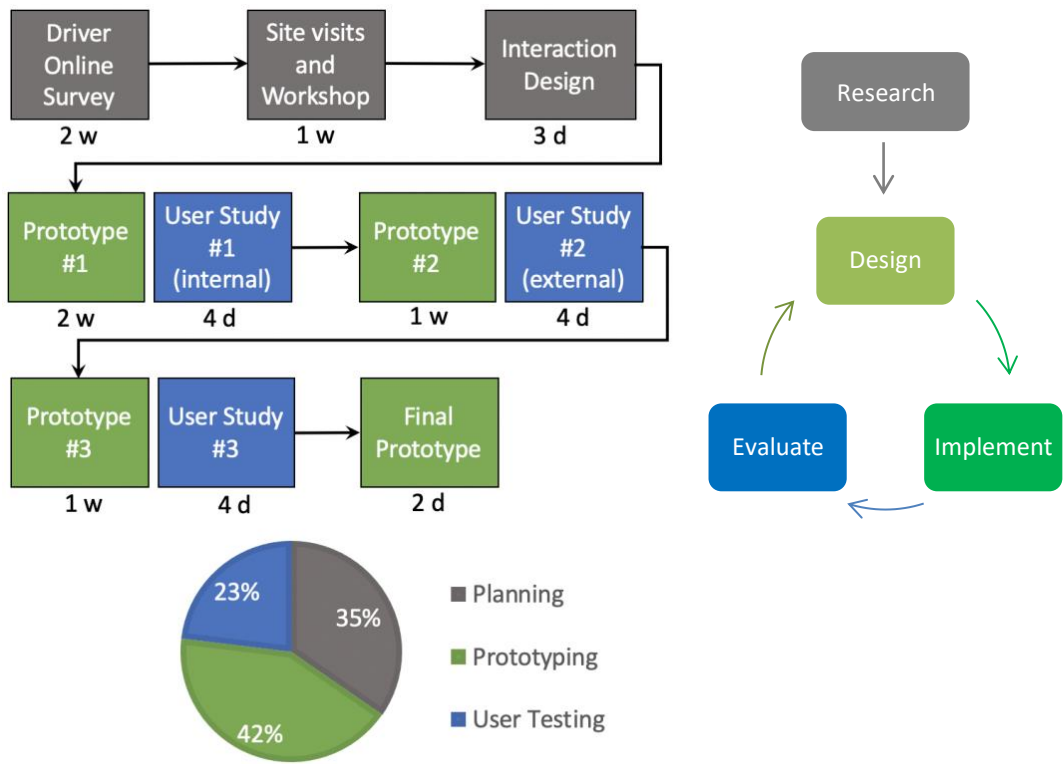

**Figure 1: Schematic of the project timeline showing the three main phases of the design process (Planning, Prototyping, and User Testing) and how much time resources they consumed relative to each other. The timeline assumes a continuous uninterrupted work process. On the right is the design process followed: research, design, implement, and evaluate.**

This paper provides answers to these questions by documenting our research and detailing our agile [35] [36] development process that is geared towards the iterative improvement of HG-controlled and mid-air haptic enriched IVISs. Specifically, and unlike other recent publications in this area, the main contribution of this work is its focus upon the design process itself summarized in Figure 1, rather than an evaluation of gesture and mid-air touch IVISs. More specifically, the paper documents our design process for prototyping a gesture-controlled automotive UI that attempts to make good use of novel mid-air haptic feedback technology. Further, the paper reflects on the challenges we faced during the prototyping and design process and provides design guidelines for future integrations of mid-air haptic feedback and HG technologies into interactive car interfaces.

The paper is structured as follows: We first give a brief background description on the two main novel technologies involved in our IVIS and how we would measure the effectiveness of our prototypes. Then, we describe the online and offline research activities that helped guide the UI and interaction design. Then, we describe the iterative prototyping and user testing process, and detail our results while also highlighting any key observations made, clarifying the logic that supported our decision-making process. Finally, we critically discuss our final prototype and give design guidelines and recommendations for the future development of mid-air haptic and HG-controlled car interfaces.

## 2 BACKGROUND AND REQUIREMENTS

### 2.1 Hand Tracking Technology

There are many off-the-shelf hand tracking technologies available on the market today. A popular choice due to its low price point and extensively documented SDK is the Leap Motion Controller (LMC) – a small USB peripheral device that uses two monochromatic infrared cameras, three infrared LEDs, and proprietary machine vision software to track the user's hands and fingers. Its field of view is 150 by 120 degrees wide, with an effective range of approximately 60 cm. In 2013, the overall average accuracy of the controller was estimated at 0.7 mm [22]. A 2017 study suggests that LMC tracking is robust and stable and errors in fingertip estimation are bounded within 4–5 mm [45]. The LMC is thus particularly good at tracking the user's palm and fingertip position, orientation, posture, and motion. Moreover, its SDK has in-built functions for simple HGs interactions making it ideal for the purposes of this study, although other off-the-shelf or custom solutions are possible. Prior to the release of the LMC development kit in 2013, Microsoft's Kinect was the *de-facto* depth-camera for HG and full-body motion tracking. Therefore, most of the relevant studies for in-car gesture interactions were benchmarked against the Kinect.

An excellent and comprehensive review on HG IVISs can be found in [23] which concluded by remarking that in 2013, HG recognition technology was not reliable enough for safety related tasks in cars, could induce fatigue with prolonged usage, and lacked immediate feedback. Nevertheless, several studies have shown that users prefer HG interactions to touchscreens in IVISs [46]. At the time of writing (February 2020), much progress has been made on the reliability of HG recognition and UI design, both through SDK updates (e.g., LMC SDK v4 was released in 2018), the use of machine learning, UI design and best practices blogs, source code published on developer websites [37], as well as many academic papers that have helped unveil and mitigate some of the major flaws found in HG-controlled IVISs (and in other mid-air spatial interactions) such as fatigue, reliability and accuracy, therefore accelerating technology adoption [38]. For instance, recent elicitation studies specific to the automotive setting [39] have demonstrated that users prefer mid-air HGs that directly manipulate a virtual object or proxy object therefore defining a holistic gesture set that is intuitive and functional for performing canonical tasks on car-dashboards while driving.

### 2.2 Mid-Air Haptic Feedback Technology

A phased array composed of hundreds of ultrasonic transducers can be used to focus ultrasound energy at a specified location in three dimensions. The spatial coordinates for creating the focal point (e.g., at the center of the user's palm or fingertip) are obtained from the LMC application programming interface (API) at 100 frames per second. Then a time-of-flight solver calculates the appropriate phase delays for each transducer in order to create a high sound pressure focus. A vibrotactile sensation is created by amplitude modulating the carrier waves on and off at 200 Hz [12] or by spatiotemporally modulating the focus along a path on the hand at different speeds and sampling rates [15].

The Ultraleap SDK contains ready-made templates for the generation of mid-air haptic sensations that are documented on the company's developer website [40] or new haptic sensations can be created using a number of primitives available to the SDK. Once a sensation is called by the API, the command is transferred from the host PC to the Ultraleap device via USB (estimated at 1 millisecond) where it is processed (1 millisecond) and converted into electrical signals (1 milliseconds) that are transduced into ultrasound waves (1 millisecond) to be emitted by the ultrasound speakers. These are then focused onto the user's palm or fingertips which are typically located 20 to 40 cm away from the device (1 millisecond). It is therefore estimated that the total latency of the system is approximately 5 milliseconds for the haptics, and 5 milliseconds for the LMC API [37].

## 2.3 Previous Mid-Air Haptic IVISs

There have been several mid-air haptic IVIS concept prototypes designed and implemented and even integrated in mock-up vehicle dashboards. One example included a custom plastic mold cast of a car dashboard with all the components assembled and fitted into the unit [20]. Two more demonstrators include the BMW and Bosch concept cars at the Consumer Electronics Show (CES) in Las Vegas in 2017. Such high-quality commercial grade demonstrators were mostly geared towards demonstrating the device hardware capabilities while also considering of the various UX requirements and even incorporating some future concept applications, such as the integration of cars into the internet of things and smart home ecosystems. The development process of these two demonstrators however is not publicly available, therefore providing little insights on the various design challenges met by the experienced teams at BMW and Bosch, nor do they share any design guidelines or lessons learnt. In stark contrast, most published academic studies such as those reported in [19] focus on very specific ergonomic and UX design aspects such as driving safety or the added value of multimodal feedback [18], and thus meticulously document all the rigorously measured human factors related to in-car HG interactions such as the task completion times and eyes-off-the-road-time, as well as their methodology, results and apparatus.

The current paper's objective is twofold. The primary objective is to document the research, design, and engineering process behind the development of a mid-air haptic gesture controlled IVIS prototype demonstrator that incorporates the most common and basic functionalities found in existing cars. The second objective is to reflect on the challenges we faced and provide design and interaction guidelines for future integrations of mid-air haptic and hand gesture technologies into IVISs [24]. To that end, we document our research methodology and design process with the intention to effectively and openly deliver UX design insights and guidelines for any future development of mid-air haptic and HG-controlled IVISs by academic researchers and industrial automotive UX/UI designers alike [25], while being transparent about the limitations of our study and recognizing that our development process was by no means free of flaws and imperfections. Thus, we henceforth choose to apply a storytelling narrative approach that can guide the iterative improvement process while supporting future researchers and designers in exploring and sharing their ideas and methods [26]. We note that in order to present all the main steps taken during the prototype development process, our narrative will often tradeoff scientific rigor for engineering and design insights.

## 2.4 Requirements for Mid-Air Haptic IVIS UI

At the beginning of the present development project, we wanted to focus on the interplay between the graphical UI, the mid-air haptic HG interactions and the IVIS functions themselves. To that end, we prescribed ourselves a set of core targets, and the HMI we would design should meet the following REQUESTs (REliable, Quick, Useful, Easy, Safe, and realisTic):

1. **Reliable** - The capabilities and limitations of the available hardware should be considered.
2. **Quick to learn** - Should only require a brief explanation of the UI before use.
3. **Useful** - The functions should match those typically found in cars today.
4. **Easy to use** - Should be logical and make sense.
5. **Safe** - Could be used without taking eyes off the road.
6. **Realistic** - Could be used effectively whilst driving.

The above REQUESTed targets were chosen since they encapsulate key specification guidelines found in best practices for general UI design [24][25] as well as specifically for IVISs [21][23][34]. It was realized from the start of the project that fulfilling individual targets or pairs of these would be relatively easy but fulfilling all of them simultaneously would be particularly challenging and required in-depth research and analysis in IVISs, plus a flexible iterative prototype-and-test design process that would allow for trial and error development.

## 3 IVIS RESEARCH

### 3.1 Driver Survey

We decided that sending an electronic survey to members of the public that drive vehicles on a regular basis would be a great way to start the development process and get input from a variety of end-user demographics. The survey questions would be designed to provide information on 1) the usage levels of the various in-car infotainment features (whilst driving), and 2) what the most common interaction methods were being used to complete typical in-car tasks.

The survey was thus created using Survey Monkey and sent out to 59 people in the USA which represents a major consumer base for most car manufacturers. Getting input from more than one market, e.g., from USA and Asia, would probably induce all sorts of cultural biases and differences that could ultimately confuse our development process. We did however want to receive feedback from different genders and ages so that we can later identify our ideal customer avatar. We therefore received answers from ages between 18 and 60 (mean age was 41) with 34 of them being female. Moreover, a screener question was used to ensure only people that drove cars completed the survey. Some of the 12 questions in the survey were multiple choice, some required a score from 1 to 9 on a Likert scale, and others required a short text input answer. The survey structure started with simple questions and followed with more complex and specific ones. Questions were clustered into groups for different topics, and answers were cross tabulated and filtered. In the case of free-text entries, we used a coding approach where we identified themes and categories and then tallied the responses that fit those categories. We therefore were able to collect both quantitative and qualitative information.

The main survey results can be compiled into a three-dimensional matrix of 1) secondary task frequency - "What tasks do you typically do whilst driving?", 2) secondary task action - "What action best describes how you complete these tasks?", 3) secondary task difficulty - "How easy or difficult are the following tasks to do whilst driving? and Why?". Out of the 10 possible secondary tasks reported in our survey, the most frequent were: Turn the radio on/off (89%), Adjust the volume (82%), Adjust the temperature (72%), and Answer incoming call (58%), while changing track or radio station, making a phone call, and setting navigation destination were less frequent. Out of the 8 actions provided, Pushing/Pressing was the main action people used for most of the tasks. Turning/Twisting was predominantly used for adjusting the temperature. Voice input was mainly used for making phone calls. Setting the navigation destination for a journey was ranked as the most difficult task because the task is generally more complex (31%) and the physical position of device is difficult to interact with (29%).

While this is not a complete representation of all the data collected, it provided us with a good indication of what to focus on when designing the UI prototype for IVISs. Designing, setting up and running the survey took 2 weeks.

### 3.2 Existing IVISs Research

In addition to knowing why and how people typically use their car's infotainment system, we wanted to know what the most common UI characteristics are currently being used in existing vehicles. Clearly the answer to this would probably display a large variation across car models and makers, however it would greatly help our development process by for example highlighting any common "must-have" features, terminology, and recent trends.

Our research methodology here included both online and offline aspects. For our own convenience, we visited car showrooms and dealerships near our office in the UK and took screenshots of online videos showcasing different existing car infotainment systems. In retrospect, we should have sent the online survey to UK recipients, and not the USA, to ensure consistency between results from the previous subsection. The majority of the cars and showrooms we sampled consisted of recent car models from almost all major manufacturers operating in the UK. The end result was a small sample of graphical UIs and IVIS menu structures that included different input methods (HGs, touch screen, physical dials). The sample was then pruned down with an aim of capturing the breadth of features and menu structures that we had sampled while avoiding repetition. Importantly, the reduced sample set included options for all the interactions and use cases that we had previously examined through the online survey. The reduced sample set was then printed and pinned to the wall (see Figure 2) to facilitate for further brainstorming during the internal interviews that followed and was also used as UI inspiration for the software designers and developers that built our IVIS prototypes.

### 3.3 Internal Interviews and Workshop

In order to take the appropriate intersection of Driver Survey results and the Database of existing IVISs, we consulted the business department at our organization consisting of sales and marketing experts. The rationale here was that human factors and UI need also to be aligned with the industry needs and business cases which are known best by business development experts rather than UI designers and UX researchers. Indeed, such an approach also termed Minimum Viable User eXperience (MVUX) aims at providing end-users with a good enough UX from an early stage of product development yet enables communication of the envisioned market value through the *a priori* gathering of meaningful business feedback and market insights [27].

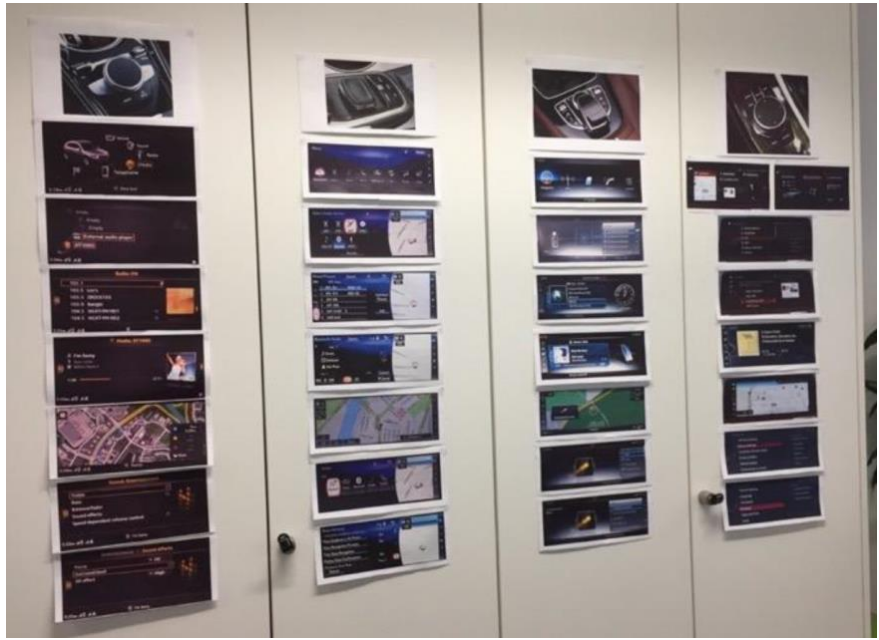

**Figure 2: Printouts of a sample selection existing IVIS input controls and menu architectures.**

The main purpose of the workshop was to agree a strategy for the development of our demonstrator IVIS prototype and decide 1) on the driving use cases to focus on, and 2) the appropriate hardware needed to run the demonstrator.

### 3.4   Convergence of IVIS Research

Once the interviews, showroom visits, online survey and desk research data were all collected, an internal workshop was organized in order to ensure a holistic approach towards meeting the six core targets was undertaken, and for us to agree on the hardware and software development strategy we would adopt going forwards. To avoid biases and ensure that we reached a more democratic result, the workshop was attended by two UX designers, two haptic engineers, and two product managers.

The workshop begun by presenting our purpose, our research findings and our pre-determined goals and requirements (the scope). We then engaged the participants in a reflective prioritization style workshop during a round table one-hour discussion. The facilitator would take notes and keep the time. After a short break, a use case matrix was created and post-it notes were used by the participants to identify and list the types of human-computer interactions involved and then suggest hardware, software, interaction design and presentation style as seen in Figure 3.

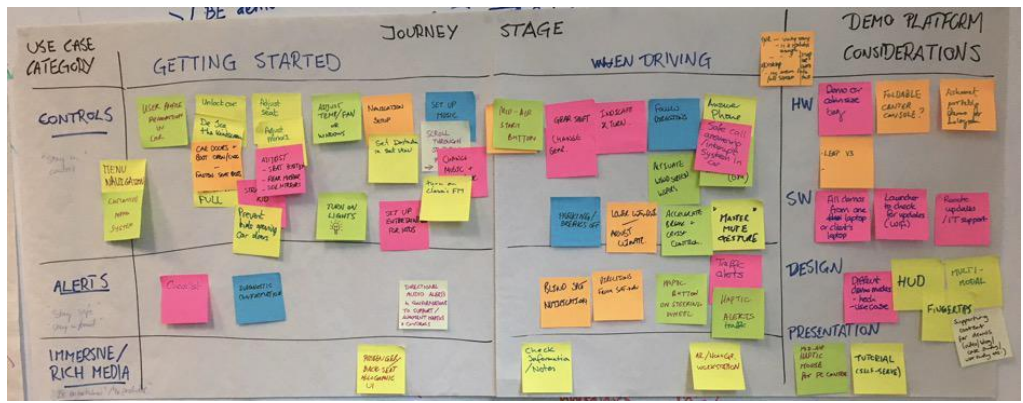

**Figure 3: Workshop ideation for use cases and interactions to be considered by our prototype.**

It was concluded that the car HMI prototype to be developed should at least cater for the following functionalities: *1)* Music control, *2)* Temperature and Fan control, *3)* Navigation map control, and *4)* Phone-call answer/reject. Moreover, it was concluded that the prototype would utilize a small LCD screen, an LMC and one Ultraleap device, both controlled by the same laptop PC therefore minimizing any hardware development requirements and allowing us to focus most of our efforts onto the HMI interaction design and testing. Importantly, the converged specification for our new prototype development was coherent with all the results derived from the online survey and the showroom visits. Finally, it was decided that the IVIS prototype should be user tested by both internal and external participants using a low fidelity driving simulator. We therefore set out to design a wireframe for the IVIS UI and define the various interactions we would test. We also begun researching for the different parts (and their associated costs) that would make up our low fidelity driving simulator within our prescribed budget.

## 4  INTERACTION DESIGN

### 4.1  Wireframes and Flows

As with most design work, we started sketching some ideas of what the UI could look like based on the existing systems we had researched that could be easily expanded into 3D and therefore be compatible with HGs and with mid-air haptics. The rationale here was that such an approach would by default ensure that core target #2 would be adequately met from the start.

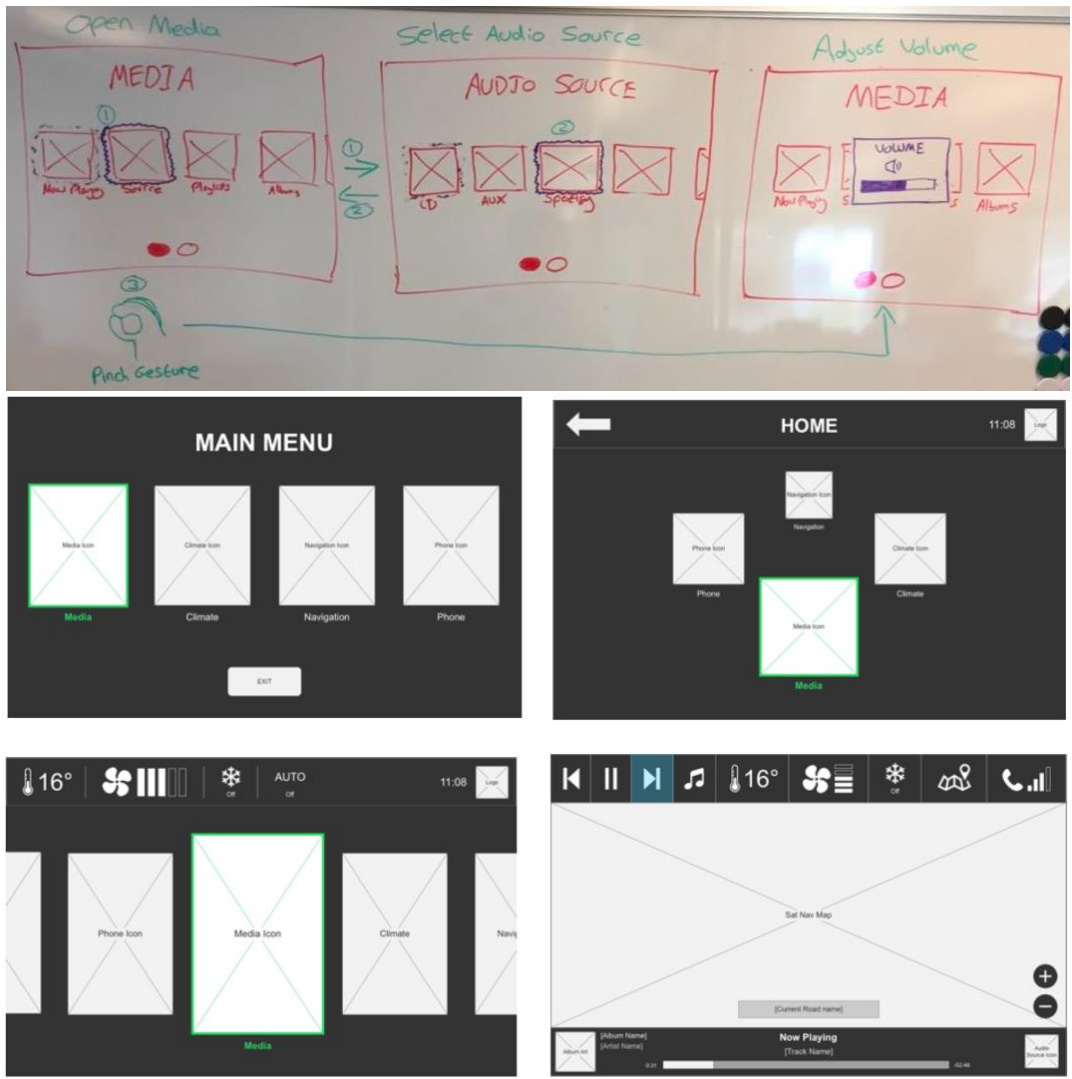

**Figure 4: Wireframe designs for the IVIS home menu.**

The first concepts designed were based on the idea of having tiles (menu items) that the user could highlight with a motion-coupled cursor that selects menu items by swiping the cursor in the desired direction. This is essentially the simplest and most obvious extension of a touchscreen style UI into the out-of-screen direction. Four example designs for the UI and the original whiteboard sketches are shown in Figure 4. The top left design is a linear menu arrangement showing four options and the exit button. It can be navigated by moving left/right through some HG (to be defined in the next subsection) which is quite easy and simple. The selected menu item is visually highlighted which we expect not to be very suitable for cars as that demands some visual attention. The bottom left and top right designs in Figure 4 are carousel variants of the previous one with the selected menu item always staying in the middle of the screen therefore probably being slightly less visually demanding. The bottom right design is of a navigation screen which has all the setting information in the top bar and the music information on the bottom bar and a map in between.

### 4.2 Interaction Style

When it comes to mid-air haptic applications, the UI designer essentially must decide between two interaction styles: *1)* Fixed-in-space (close to reality, physics based), or *2)* Gesture based (specific hand poses and movements). Fixed-in-space mid-air haptic interactions are more intuitive, and easy-to-learn since they more closely resemble those of existing touchscreen interfaces. Gesture based interactions on the other hand can be operated without visual aids, are not fixed in space and so can be performed more naturally and with less stringent accuracy requirements. We therefore decided that gesture-based interactions were the most appropriate for an IVIS, because it meant the driver could operate it without taking their eyes off the road or wasting time 'feeling around' for invisible mid-air haptic buttons. This view was also supported by both previously carried out studies in Glasgow [18] and in Nottingham [19]. Core target #4 would therefore have to be addressed through a simple and intuitive UI design rather than relying on pre-learned and already familiar interfaces.

### 4.3 Mid-Air Haptic Gestures

One of the biggest challenges when interacting through HG-based interfaces is needing to remember all the HGs required to operate the application. One would say that it's a bit like learning a new language; a sign language in this case. For this reason, it was important to keep the number of gestures to a minimum [28], whilst still ensuring the user can navigate and use the interface effectively. Ideally, a single HG would be used for many different actions, depending on the context. Moreover, these HGs would always have a similar and conceptually intuitive action assigned to them. We therefore aimed to take inspiration from [9] [29] and [33] and also following the classifications of HGs provided by Pickering et al. [41] while recognizing that some of the HGs would require some guidance or explanation before first use.

Choosing the accompanying mid-air haptic feedback was going to be the next challenge. Through various reports taken from the literature [9] [18] [19] [20] and in-house developed experience by our haptic engineers, we have noticed that some haptic sensations are more reliable than others. For example, it was reported in [19] that users preferred haptic sensations that where dynamic, i.e., would change according to changes within the IVIS. Similarly, it was reported in [18], that users preferred haptic sensations that would give a clear confirmation feedback to the user after an action is performed, e.g., the pressing of a holographic button.

The wireframe designs shown in Figure 4 and a list of requirements and guidelines for HGs and corresponding haptics shown in Figure 5 were then assembled into a coherent report and communicated to a development team that would then produce a Unity build (a cross-platform game engine written in C++ with a C# scripting API) of the desired IVIS prototype. This preparatory step took approximately 3 days of design work.

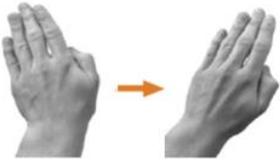

Figure 5: List of the Prototype #1 HGs and mid-air haptics. Some of these have changed in Prototype iterations #2 and #3 as noted in the text. The HGs are presented using the left hand assuming the driver seat is on the right of the car (e.g., for UK, India, and Australia etc.). These can be reversed for use in cars with the driver on the left side of the car.

# 5 PROTOTYPING AND USER TESTING

## 5.1 Prototype #1

Figure 5 shows the gestures and corresponding mid-air haptics that were used in the first prototype we developed. These five HGs chosen were inspired from recent relevant literature [9] [18][19] [20] to allow for direct comparison and also since these have already undergone user testing and have emerged through user feedback elicitation [39] or online survey studies tested via agreement metrics propped in [42]. A swiping HG was associated with moving to the next menu navigation as described in [33]. This gesture is part of the LMC SDK gesture class and so was easily coded into the prototype application. A hand twist gesture was used to navigate to the previous menu item. Here, the palm of the hand would twist from being downward facing to being upward facing, and then back to downward facing. This gesture is not available from the LMC gesture class and so had to be programmed manually. This was done by tracking the palm vector direction and triggering the gesture action when a quick and big change in its z-axis (i.e., vertical) direction was detected. A pinch and move up/down HG was used for a more accurate control over a system function such as volume and fan speed level. The pinch gesture was part of the LMC SDK gesture set. Within the prototype, we allowed the user to release the pinch at any height, and then re-pinch at any other height without resetting the volume level. A hand tap gesture was implemented to indicate a selection or confirmation action. This gesture was also not available through the LMC SDK and so we had to manually create it. The LMC API can report hand position, speed, and acceleration. We therefore used the latter to detect a rapid change in hand speed in the z-direction. Finally, grab and release HG was used to reset the navigation menu to the home screen. Grab interactions were also part of the LMC SDK. Other HGs could have been included however for the sake of simplicity it was decided not to since these five satisfy the minimum required specifications that were decided at the internal interviews and workshop.

The accompanying mid-air haptics are shown in the second column of Figure 5. At this initial prototyping stage, we implemented just one haptic sensation, namely a circle haptic, but presented in five different ways. We made this choice so that the haptic feedback would be consistent, but also different between HGs. During a swipe HG, a circle haptic would be presented to the middle of the palm for a period of 0.5 seconds thus creating a short but noticeable tap sensation. The circle had a diameter of 4 cm. A modulation frequency of 70 Hz was used to render circle haptic as recommended by ref [15]. For a hand twist HG, a similar circle haptic but now split into two consecutive taps was presented at the end of the HG as to communicate a confirmation of the gesture and also to use the often used phrasal verb of "to double back" which means to go back in the direction one has come from. The duration of this sensation was shorter at 0.3 seconds per tap with a 0.1 second gap between them. For a pinch and move HG, a similar circle haptic was presented through the gesture starting from the moment of the user forms a pinch, until the pinch is released, as to communicate to the user that she is currently "in control" of the volume proxy object and that the IVIS is responsive and follows her actions. For the hand tap gesture, a dynamic circle haptic sensation was designed. The circle radius would start small at just 1 cm and increase to 3 cm linearly during a time interval of 0.7 seconds as to communicate to the user that her action was "opening" or "enlarging". Finally, for the grab-release HG, the reverse haptic sensation was used, that is, a circle haptic that started big and shrank to 1 cm was used to indicate to the user that she was "closing" the applications and going back to home. The precise timings of these haptics effects were designed empirically by our haptic engineer and iteratively refined with a UX designer from our team but were not individually user tested, as that would consume much more of our time and is a separate study in itself.

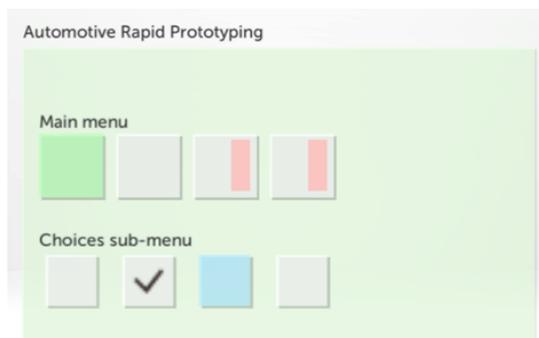

**Figure 6: Example graphics for Prototype #1.**

To test the performance of the interactions shown in Figure 5, a simplified IVIS prototype was assembled. This UI was deliberately basic and graphically unskinned and made of single-color block buttons (see Figure 6 and 7). Despite this simplicity, the prototype sufficiently represented the UI and information architecture (IA) for a round of internal user testing. The development, testing and final integration and of the HG input set and their corresponding haptic feedback effects into a Unity project took approximately 2 weeks of software development time.

## 5.2 User Testing – Round #1 (Internal)

Seven colleagues of ours (3 female) aged 27 to 43 (mean age 35) volunteered to participate in the internal user testing pilot. All participant volunteers had a UK driving license and were familiar with the technologies employed and were also shown what the HG and mid-air haptic set of Figure 5 *a priori*. Participants were then asked to perform a set of secondary tasks using our developed prototype whilst operating a low fidelity driving simulator.

The simulator we put together was by no means state-of-the-art, however adequately and cost-effectively served the purpose of iterative testing and development of our prototypes. It consisted of a GT Omega Racing cockpit (GTO-003RS), a Logitech 920 Steering Wheel and Pedals, a Windows PC, 3x Dell widescreen 24-inch Monitors, a small 11-inch monitor was used to display the UI. The Carnetsoft Driving Simulator for training and research was used while driving on a simulated motorway. Finally, an Ultraleap STRATOS Explore Dev kit equipped with a LMC was used for recognizing HGs and returning mid-air haptic feedback onto the user's left hand since we assumed that the driver is sat on the right of the car as is the case in the UK. Assembling and testing all parts used by the simulator took one full day.

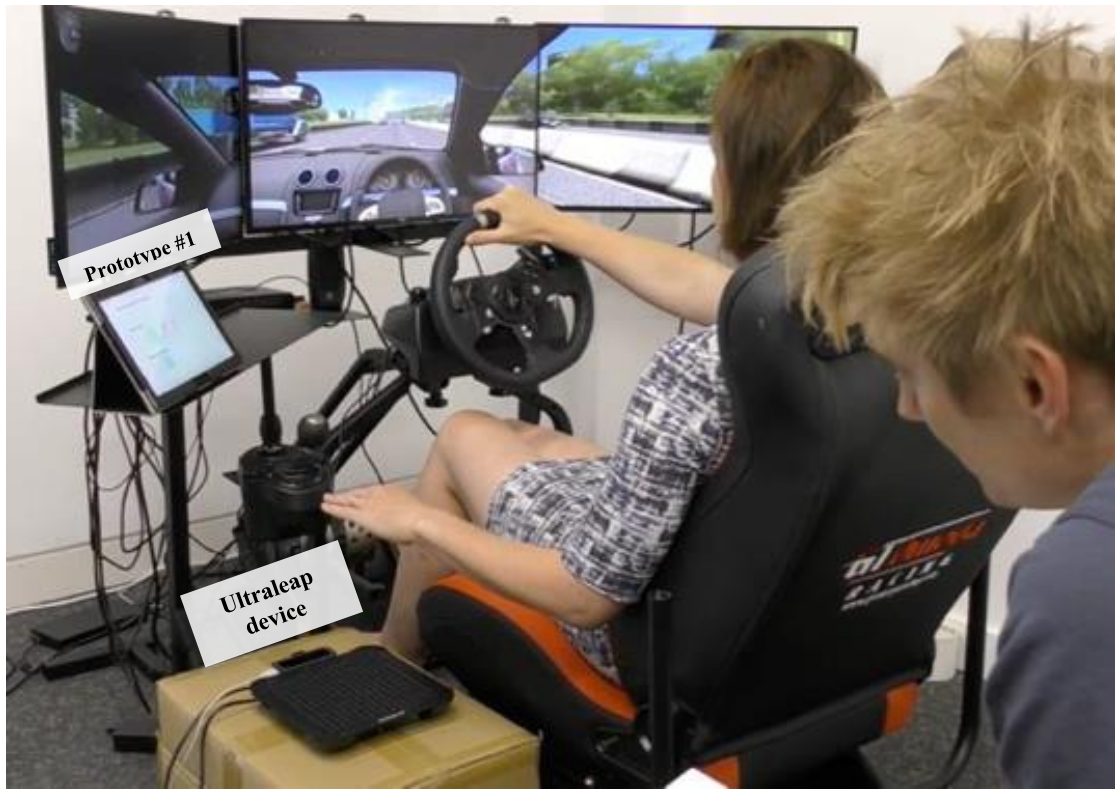

**Figure 7: Low fidelity driving simulator with mid-air haptic and HG input control capabilities. Prototype #1.**

While driving on a simulated motorway, each participant would first navigate through the menu items using swipe and hand twist HGs, select a menu item using the hand tap HG, adjust a value using the pinch and move HG, and then go back to the home screen using the grab-release HG. After repeating these actions a few times, each participant was asked to accept or reject an incoming call using the hand tap or grab-release HG. Each participant test took about 30 minutes. At this stage, our primary aim was to find out if the proposed interactions were feasible, reliable and practical to use, and if there were any obvious pitfalls or problems to be reported. Due to the small participant group size, we could assess the performance of the prototype to that respect through direct observation and by talking to the participants and taking notes of their honest feedback. No quantitative data was taken at this stage. Participant user testing of Prototype #1 took three full days and one more day for assembling and processing all the collected data.

Following the first round of user testing, it was quickly realized that HG tracking reliability was the biggest concern, especially with the swipe gesture. Users would swipe in many different ways and speeds and the optical camera would fail to recognize the HG reliably or would record false positives. Participants expressed frustration to that effect and were even able to describe why the system was misbehaving. Specifically, the swipe HG is not a discrete one, with a clear starting and end point. Therefore, the system would either fail to detect it in its entirety or would falsely detect a swipe when there wasn't one.

At the end of each user trial, we specifically asked participants to tell us if the proposed gestures were "acceptable" to them, and if in their opinion and after some refinement would potentially satisfy the REQUESTed targets. User acceptance testing (UAT) is commonly used during the development of information systems. In our case, the objective of UAT was to assess and confirm that the system is usable from an end user perspective (i.e., operational ease-of-use) [44]. Thus, participants could reject a proposed HG and could also comment on any shortcomings with respect to meeting the 6 targets of the IVIS.

All HGs were accepted by our pilot study as input control methods (i.e., scored higher than 4/7), but some participants thought the swipe gesture was potentially physically tiring to perform repeatedly [23]. All haptic effects were also accepted except for 'Next' and 'Previous' menu navigations which scored just 3/7, i.e., only three out of seven participants said they were acceptable haptic effects. Therefore, following the recommendations from the study participants, these were changed in the second prototype iteration from single and double taps to a hand-scan sensation which is basically a haptic vertical line that moves across the user's hand in the direction the swipe or twist and feels like a hand-scan. The line therefore moves from left to right for a swipe gesture that moves from left to right, or from right to left for a hand twist. The updated haptic is shown in the second column of Figure 8. Changing the haptic output was took an additional day of development and testing.

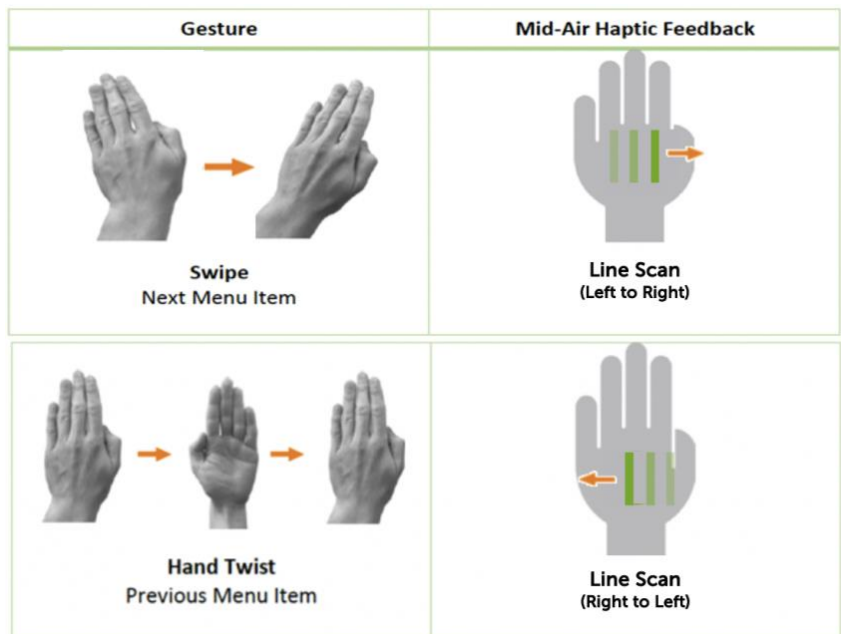

**Figure 8: Revised Haptics for Next and Previous HGs used in Prototype #2.**

## 5.3 Prototype #2

In the second prototype iteration, the graphics of the IVIS were significantly improved and polished up as to look more like a commercial demonstrator using the wireframes shown in Figure 4 and inspired by the UIs collected during our showroom visits and online background research. The accepted interactions and haptics were carried over from Prototype #1. Namely, the user now had to perform a Swipe gesture to move onto the next menu item, a Hand-twist to move back to the previous menu item, Hand Tap to select an item, and a Grab-release gesture took the user back to Home Screen. Pinch and move up or down gestures were used to adjust volume, temperature and fan speed when those features were highlighted. Keyboard shortcuts were used to trigger incoming phone calls and modals that suggested a destination. Users would accept them with Hand Tap gesture or reject them with the Grab-release gesture. To mitigate the reliability problem of the Swipe HG in the Prototype #2 we restricted its trigger to within a bounding box interaction region. Namely, a second condition was added to the Unity application requiring that for any HG input to be triggered, it had to take place within a cube of dimensions 30 x 30 x 40 cm located 5 cm directly above the Ultraleap device as shown in Figure 9. The bounding box for the hand gestures interactions was indicated to the users during the 15-minute familiarization stage by the host. Further calibration of the dimensions of this interaction region warrants a separate study altogether and would depend on the exact ergonomics and relative positions of the car dashboard, IVIS, and potentially the car seat and user preferences. For the purpose of this work, we considered that an interaction volume of 36000 cubic centimeters was sufficient. The development, integration and testing of the revised Unity project took approximately one additional week of software development time.

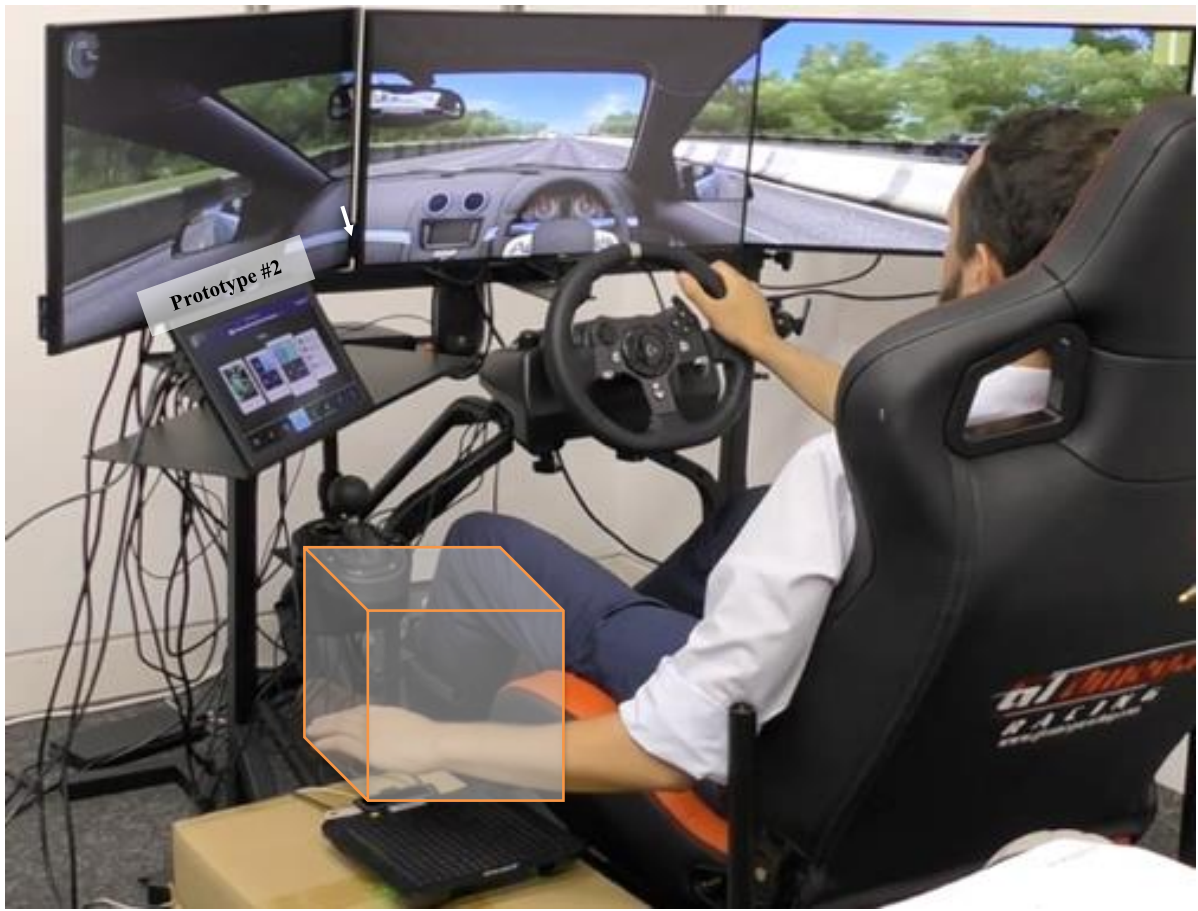

**Figure 9: Prototype #2. Low fidelity driving simulator with mid-air haptic and HG input control capabilities also showing the bounding box for HG interactions.**

## 5.4 User Testing – Round #2 (External)

Seven external participants (2 female) were involved in this round of user testing. All were aged between 23 and 43 years old (mean age 29), right-handed, and had a UK valid driving license. None of the participants were familiar with the technologies involved and so a 15-minute familiarization phase was planned so that they could understand the basics of HGs, optical hand tracking, and ultrasound mid-air haptics. The participants were rewarded with a £20 Amazon voucher. The study procedure involved a short pre- and post- study verbal questionnaire, as well as a live interview during the driving simulation. The study was approved by our ethics committee.

During the study, the participants would sit in the cockpit shown in Figure 9 and the interviewer would ask them about their first impressions and expectations such as "what do you think these bars under the three icons represent?" and "how do you think you can interact with this menu using mid-air gestures?". These questions allowed us to get unbiased and fresh insights about the user's first impressions. Then the interviewer would explain what the intended HG interaction should be and invite the participant to try themselves with prompts like "try highlighting the …", "try selecting the …" and then asking for feedback "how did you find the…?", and "was that an appropriate gesture for doing …?". The interview would be as fluid as necessary yet as structured as possible and keeping to a time limit of 45 minutes while also going through the set of all interactions in this sequence: Next, Previous, Select, Back to Home, Increase/Decrease, Navigation Tasks, Phone Tasks. For each interaction, the interviewer would ask between 8 to 12 questions and record the key points raised relating to things that worked well or did not and therefore invited closer review by the design and development teams. Then, the participant would be tasked with driving an automatic vehicle on a motorway using the simulator while also performing a sequence of approximately 14 secondary tasks verbally instructed by the interviewer, e.g., "skip to the next track", "answer the phone", "end phone call", etc. Finally, the participant would stop driving, and would be asked a few generic questions like "were there any tasks you struggled with?". The key points from the recordings were then noted for post processing and analysis and used for the $3_{rd}$ prototype iteration. Participant user testing of Prototype #2 took three full days and one more day for assembling and processing all the collected data.

Several qualitative results and observations were made during this study. The most important one was that the swiping HG was still not working reliably and was also tiring to perform repeatedly. This also resulted in participants having to glance at the screen in order to affirm that they had highlighted the correct menu item. We therefore dropped the swiping HG for menu navigation from Prototype #3 but kept it for the skipping track task since this was performed less frequently according to our online survey and is associated with audible feedback (new music track starts playing) rather than visual (new menu item is graphically highlighted). Similarly, the hand twist gesture was reassigned to going back a music track because it was reliable and widely accepted as a HG for going back.

The mid-air pinch gesture and corresponding haptics were widely accepted by the participants for adjusting settings such as temperature and fan speed. Participants reported that it was reliable, intuitive and easy to perform. Finally, the hand tap gesture and corresponding mid-air haptics was also widely accepted for selecting items and answering calls.

## 5.5 Prototype #3

After two prototyping and testing iterations, we had mostly reached a conclusion on the gestures and haptics associated to the different IVIS functions. We had also concluded on the UI layout, its content, its visuals, the IA, and what functionalities would be enabled through mid-air interactions and how mid-air haptic feedback would be applied to improve UX and reliability of the HGs. However, having dropped the swipe HG from menu navigation, a new navigation method was now required as well as a new menu wireframe. It was also decided that there should only be four selectable modes (media, temperature, fan speed and navigation) and no submenus to simplify the IA even further. This was quite a large change to make at this late stage of the prototyping process, however that is what the user testing was pointing towards. Two candidate interactions were therefore proposed and tested.

### 5.5.1 Finger Pose Menu Navigation

Finger pose gestures were previously studied in [18] and were found to perform well in automotive settings. With this approach, the user can select one out of four menu items (1) Media, 2) Temperature, 3) Fan Speed, and 4) Navigation), by placing their hand and fingers in a certain way. In this case, it is the number of fingers shown to the LMC makes the selection (see Figure 10). Finger-counting, also known as dactylonomy comes naturally to most people and can serve as a form of manual communication in marketplace trading or SCUBA diving (hand signaling) and also in games such as Morra (an ancient Roman and Greek times) although it does have some variability between cultures and how they represent different numbers. In our implementation, we adopted the North American and UK system where the index finger represents 1, the index and middle fingers represents 2, and so on until the index, middle, ring, and little fingers represents 4. The LMC API has an in-built function for counting fingers which is quite reliable and easy to implement. Once the system recognizes the finger pose, the user feels a haptic line effect (known as 'finger scan') move across their fingers. It is not however clear what would be a good way forward in the case of more than 5 menu items therefore restricting the IA.

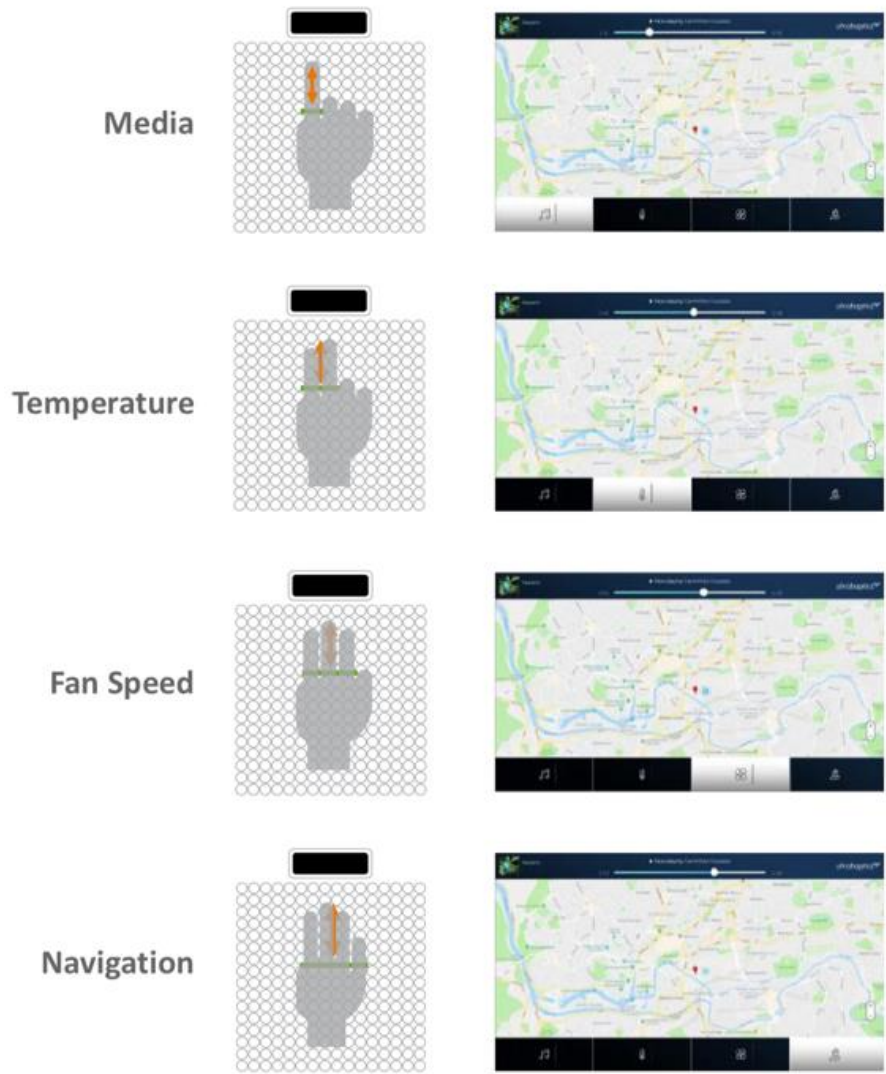

**Figure 10: Finger pose and mid-air haptics menu navigation.**

*5.5.2   3D Radial Menu Navigation*

With mid-air interactions, there is an extra spatial dimension to work with. HG input is free to move in 3D space, rather than the existing 2D surface interactions of touchscreens and traditional physical buttons and dials. Further, most people have very good proprioceptive awareness related to the motion of their arms without having to look. For example, one can adequately throw a tennis ball in any desired direction with their eyes closed [32]. With this human capability in mind, we designed a 3D radial menu gesture navigation method (see Figure 11) consisting of four options that also appear on the screen: 1) Media, 2) Temperature, 3) Fan Speed, and 4) Navigation, positioned in a West, North, South, and East arrangement, respectively. The Pinch HG (similar to the universal symbol for "OK") activates the 3D menu. While holding the pinch gesture, the user can move their hand in these four directions in a similar manner as moving the gear stick forwards, backwards, left or right. Releasing the pinch will make the menu selection and an expanding circle haptic effect (aka 'Open sensation') similar to the Hand Tap is projected to the palm to confirm to the user the selection. This 3D radial menu approach is actually quite novel and has thus not been studied before, at least not within an IVIS context.

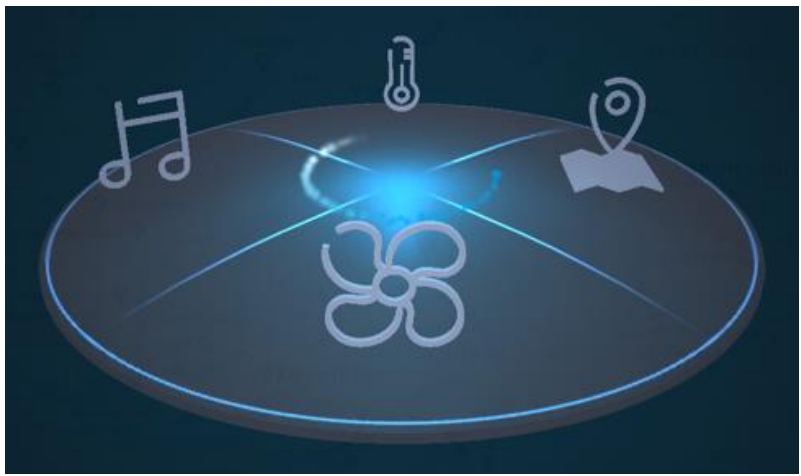

**Figure 11: 3D Radial menu screen shot.**

Menu navigation is an essential part of any IVIS and thus changing the interaction technique after two prototype iterations is certainly a major one. Both proposed navigation methods were implemented in the third and final prototype that could be switched by pressing a keyboard shortcut. Therefore, the prototype would be demonstrated with either the Finger Pose or the 3D Radial menu for navigation, while the remaining sub-menu items and functional controls remained common. The development, integration and testing of the revised Unity project for Prototype #3 took approximately one week of software development time.

Hence the third and final prototype consisted of the following functions and just 3 core HGs in total as to keep things simple and in agreement with the recommendations of [28]. Swiping and Hand Twist are not considered core gestures as they are only used for one function within the media mode.

- o Finger Pose menu navigation: *1-4 fingers (see Figure 10)*
- o 3D Radial menu navigation: *Pinch and Move left/right/forward/backward (see Figure 11)*
  - • Adjusting volume / temperature / fan: *Pinch and Move up/down*
  - • Skipping a track: *Hand Swipe (Media mode only)*
  - • Previous track: *Hand Twist (Media mode only)*
  - • Pausing / Playing media: *Hand Tap*
  - • Answering a phone call: *Hand Tap*
  - • Declining a phone call: *Grab-Release*
  - • Zooming in / out on the SatNav route: *Pinch and Move*
  - • Confirming a new navigation route: *Hand Tap*
  - • Declining new route: *Grab-Release*

### 5.5.3 Additional Auditory Feedback

Auditory feedback in driving environments can reduce looking away time and if presented together with to mid-air haptic HGs it can reduce eyes-off-the-road time without negatively impacting the driving performance [29] [30] [18] nor mental demand [7]. We therefore decided to incorporate auditory feedback to our IVIS prototype. For the 3D radial menu navigation method, verbal confirmation was given while highlighting one of the four options: "Media, Temperature, Fan, Navigation" followed by a short "ding" sound to indicate the selection of a menu item (in addition to haptic feedback) when the pinch gesture is released. For the Finger pose menu navigation method, verbal confirmation was used to indicate the selection of one of the four options: "Media, Temperature, Fan, Navigation", without any "ding" sound. Finally, verbal confirmation is given when accepting/rejecting a new route or call.

### 5.6 User Testing – Round #3 (External)

A third and final round of user testing was performed to evaluate the effectiveness of the new changes and identify any further unexpected shortcomings. Eight new external participants (3 female) were involved in this round of user testing. All were aged between 25 and 45 years old (mean age 32) and had a UK valid driving license. The participants were rewarded with a £20 Amazon voucher. The study protocol was essentially the same as for round 2, but with particular focus given to the testing of the 3D radial menu and finger pose menu navigation methods. We also removed some tasks from the script that covered parts that hadn't changed since the last round so that we could focus more time and attention on the new parts. Participant user testing of Prototype #3 took three full days and one more day for assembling and processing all the collected data.

When participants were asked which HG they would instinctively use to skip a track (before using the demo during the first impressions phase), the majority (six participants) performed a swiping motion with their hand. When it came to performing the skipping track task, participants were therefore very accepting of the gesture and haptic feedback. Hand Tap gesture for playing/pausing music was also widely accepted with five out of the eight participants saying that they liked this HG and found it easy to perform. Finally, only one in eight preferred the 3D radial menu whilst using the driving simulator, therefore declaring the finger pose navigation method as the clear winner in terms of UX. One potential reason for this is that the Finger menu selection method is a more direct and much simpler (one-step) action compared to the 3D radial navigation.

## 5.7 Final Touches

After three iterations of prototyping and user testing our design and development project had converged to a well-functioning prototype that seemed to sufficiently meet the REQUESTed targets. While further iterations could further improve our prototype, eminent deadlines and budget restrictions signaled the end of this project. Following the last user testing iteration, some final additions and changes were made as finishing touches to the prototype. These included:

- Graphically fade out parts of the menu that are not being used. This can help users concentrate on the secondary task without the addition of unnecessary visual clutter.
- Add a constant but low intensity haptic circle sensation fixed 20cm above the device as to help users initiate their interactions from an optimal location.
- Disable the detection of certain gestures for 2 seconds after other gestures are performed. This led to less HGs being triggered accidentally in sequence.

The development and testing of the final prototype took approximately two more days of software development time. The complete timeline of the project is simplified and presented in Figure 1. We note that the split between planning, prototyping and user testing is almost equal, but appears to be slightly lagging in user testing time. However, it is worth noting that user testing relies on the availability of external participants to come to the testing site and therefore this phase actually took a lot more time that indicated by Figure 1.

## 6 DISCUSSION

Our study is the first to develop, test, and document the design and engineering process of an IVIS that is HG-controlled and has mid-air haptic feedback. As such, it is important to note that in order to cover all this ground we had to sacrifice and trade-off study rigor for speed and flexibility during prototype iterations. For instance, very little quantitative data was generated in favor of more qualitative user feedback that could then be easily communicated back to designers and developers who could practically iterate the three versions of the prototypes. Having said that, it is paramount that in the future we revisit the individual assumptions and decisions made and study these with the appropriate rigor. Namely, it must be both appreciated and stressed to the reader that further quantitative studies are needed to truly assess and therefore improve the performance of the designed IVIS prototyped herein, and especially aspects relating to safety in increasingly complex driving scenarios and realistic simulators. For now, however, a qualitative assessment will suffice by simply reflecting on the original REQUEST (REliable, Quick, Useful, Easy, Safe, and realisTic) targets set out at the beginning of the study:

**Reliable:** HG interfaces are currently limited by the capabilities of the tracking system used and how these are implemented within the application. While the LMC is much more reliable and easier to use than Kinect, it remains a development kit that is not out-of-the-box optimized towards HGs to be used in cars while driving. Through our testing and experimentation, we have strategically aimed to identify and use the most reliable gestures and avoided using gesture sets that the LMC is known to struggle with or which interfere with driving and can cause fatigue when used frequently. Further, we have introduced additional conditions such as an interaction bounding box (see Figure 9) intended to reduce the LMC false detection of unintended HGs thus improving reliability. Finally, we have also tried to accompany HGs with mid-air haptics that convey contextual or subconscious information through the tactile (invisible) information channel. While rigorous testing is already underway by several groups aiming to help identify and quantitatively measure the reliability of HG controlled IVISs [39] [9], we think that these do not yet address or adequately utilize the full capability potential of mid-air haptics towards improving the reliability of HG-controlled interactive IVISs. To that end, we strongly believe that proper investigations that couple interaction design, hand tracking and mid-air haptic capabilities are needed if we are to unveil deep synergies between these technologies and thus have a step change in the reliability of such integrated systems.

**Quick to learn & Easy to use:** There are only three core HGs in our prototype for the user to remember for controlling nearly all functionality: (Pinch & Move, Hand-Tap, Grab-release). The core gestures are used for different purposes depending on the context. The idea being that users will associate a certain gesture with positive actions (or "yes"), another with negative actions (or "no"), and another with adjusting values. Accordingly, the UI visuals have been designed to feel familiar to users that are being used in modern in-car infotainment systems such as the ones we surveyed from online research and by visiting different car showroom and dealerships in the UK. Therefore, the purpose of all the UI features and their icons were intuitively understood and anticipated by nearly all 15 external user testing participants.

**Useful:** Features in the final prototype include:
- Media – Controlling media by adjusting volume, skipping tracks, playing/pausing etc.
- Climate Control – Adjusting the cabin temperature between 16°C and 26°C
- Fan Speed – Adjusting the rate that air is circulated around the cabin
- Navigation – Exploring the SatNav map and accepting suggested routes
- Communication – Answering, ending and declining incoming phone calls

These features are commonly used in cars today (or in smart phones connected to the car). Users of the developed prototype should be able to relate to them and understand why interacting and controlling these through mid-air HGs whilst driving is useful. Moreover, these controls have been reported in our survey as the ones performed most often.

**Safe:** All the interactions have been designed so that the user does not need to be precise in their hand placement or arm movement to operate the application effectively. The mid-air haptic and audio feedback also confirms the driver's commands, thus reducing the need to glance at the screen to check. In addition, the amount of movement required to perform tasks has been strategically kept to a minimum, and circular HG movements have been omitted entirely to reduce cognitive load and unnatural motions.

**Realistic:** After the results of the third round of user testing, we concluded that users were able to use the application prototype to complete typical infotainment tasks whilst using a low fidelity driving simulator. Notably, this research paper does not claim to prove the system's viability or to completely overcome long standing problems related to fatigue. It does however suggest that there is real potential in the proposed IVIS technology and UI design. This is partly because the prototypes have been designed to be backward compatible with touchscreens and physical dial controllers. Indeed, it is unrealistic to assume that mid-air haptic gestures will always be the best or preferred way of inputting commands and information into the car system. For example, when the car is stationary, using a touch screen to select a feature may be faster than using mid-air gestures. Similarly, voice input might be a much more natural communications medium for road navigation [31]. Nevertheless, HGs with mid-air haptics already show clear benefits in terms of user preference, improved accuracy, and reduced driver distraction [19] that can be further improved through the inclusion of multimodal (especially audio) feedback [18]. In-car demos and testing of both hand tracking and mid-air haptic technologies are yet to be performed on real roads. Such studies require larger budgets, insurance, safety, ethics and license considerations that are sure to follow and will offer further insights to the challenges ahead.

## 6.1 Further Tips and Hints

Following all our background research, participant feedback and prototype iterations described herein, we have compiled a list of recommendations to guide future IVIS design thinking processes. These guidelines are by no means universal however may in some cases also be transferable to other interactive mid-air interfaces:

### 6.1.1 Make gestures multi-purpose

Don't try assigning each command its own gesture. Users will not remember them all. Instead, have the same gesture do different things depending on the user's context within an application.

*6.1.2 Focus on use cases where mid-air interaction is beneficial*

Avoid designing for use cases where a touch screen, physical dials or voice control would almost always be more appropriate, such as inputting an address.

*6.1.3 Avoid swiping gestures for navigating*

People perform swipe gestures in so many different ways and directions. Motion cameras rarely detect them properly. They also become physically tiring if repeated many times.

*6.1.4 Opt for controls that come to the user*

One of the advantages of mid-air haptic controls is that users do not have to reach for the controls. Instead controls can attach to a user's hand upon performing a gesture anywhere the motion capture device can see them.

*6.1.5 Use clear, simple and strong haptics*

Complicated ultrasonic haptic effects are not suitable for the simple and intuitive haptic feedback needed in an IVIS. Besides, most people will not notice the subtle differences between unique haptic effects. Simple shapes such as circles and lines are therefore more than sufficient.

*6.1.6 Link haptic properties to the values being adjusted*

For example, as the volume increases, increase the strength of the haptic feedback. This helps the user realize the result of their gesture input without needing to glance at the screen but instead through subtle tactile cues.

*6.1.7 Prototype gestures before committing to graphics and an IA*

When it comes to mid-air interfaces, the gestures are directly linked to the structure and visual layout of the application. This is an important realization that also hindered us from taking existing UI and simply adding haptics to it. Designing the IA of a 3D interface is therefore a research project in its own right and the present study has likely only explored the tip of this iceberg.

*6.1.8 Use audio feedback as well (where appropriate)*

Whilst it is best not to depend exclusively on audio feedback for an interface, it can be a worthwhile addition to alleviate dependence on visual feedback. This can make the system much safer to use whilst driving.

# 7 CONCLUSION

In this paper, we have documented the design process for the prototyping of a hand-gesture (HG) controlled automotive user interface (UI) that also encompasses novel mid-air haptic feedback technology. The design process (see Figure 1) included an online survey, business development insights, background research, and an agile framework component with three iterations of prototyping and user testing feedback loops performed with the aid of a low fidelity driving simulator. The resulting prototype integrates and improves on many results and recommendations found in the literature to produce a mid-air haptic HG-controlled automotive UI.

The developed prototype of an in-vehicle infotainment system (IVIS) converged to having just four main control functions: 1) Media, 2) Temperature, 3) Fan Speed, and 4) Navigation. These were deemed to be the most popular functions found in existing IVIS and used the most by our sample online survey. Different HG menu navigation methods were investigated. A one-hand finger counting HG was the most preferred one. Also, after experimenting with various HGs for the operation and control of the four IVIS functions we converged to just three core HGs (*Pinch & Move, Hand-Tap and Grab-release*) that are used invariably to perform secondary tasks, e.g., adjusting the temperature, pausing music tracks, answering phone calls.

All HGs were "haptified" with the aid of ultrasonic mid-air haptics, a technology that accurately focuses ultrasound waves onto the user's palm or fingers to provide instantaneous or continuous tactile feedback thereby confirming a HG input action or informing the user about the presence or state of a virtual button or dial. The implementation of effective mid-air haptic feedback was a key and equally important interaction design consideration during the agile development process of the IVIS prototype. This ensured that the performance of different HGs in terms of our six core UI REQUESTed targets (REliable, Quick, Useful, Easy, Safe, and realisTic) were heuristically and iteratively optimized. Further and more rigorous testing is however needed to ascertain the performance of these mid-air haptified HGs and investigate their generalizability to other IVISs. Therefore, future work should, *inter alia*, seek to evaluate and further optimize the technology and its implementation within a more real-world driving setting.

Finally, the storytelling narrative of this paper aimed to describe the various challenges we faced during the design process and how some of these were overcome. We hope that through this design diary on mid-air haptics and hand gesture controlled interfaces for cars we can bridge the gap between commercial demonstrators that only present a final and well-polished product but do not detail their design process, and rigorous academic research that meticulously detail their methodology and results but focus only on isolated parts of often more complex interactive systems.